\documentclass[11pt]{article}
\usepackage{array}
\usepackage{chngcntr}
\usepackage{footnote}
\usepackage{natbib}
\usepackage{bm}
\usepackage{setspace}
\usepackage{amsmath,amssymb}
\usepackage{dsfont}
\usepackage{subfigure}
\usepackage{graphics}
\usepackage{epsfig}
\usepackage{multirow}
\usepackage{color}
\usepackage{times}
\usepackage{threeparttable}
\usepackage[]{hyperref}  

\usepackage{amsthm}
\numberwithin{equation}{section}
\newtheorem{theorem}{Theorem}[section]

\newtheorem{proposition}{Proposition}
\theoremstyle{definition}

\newtheorem{remark}{Remark}[section]
\newtheorem{algorithm}{{Algorithm}}

\renewcommand{\thefootnote}{\arabic{footnote}}
\usepackage{amsthm}

\usepackage{amsmath}
\usepackage{amssymb}
\usepackage{amsfonts}
\usepackage{multirow}
\usepackage{amsthm}
\usepackage{algorithmic}
\usepackage{algorithm}
\usepackage{times}
\usepackage{bm}
\usepackage{natbib}
\usepackage{graphics}
\usepackage{booktabs}
\usepackage{graphicx}
\usepackage{tikz}
\usetikzlibrary{decorations.shapes}
\usetikzlibrary{plotmarks}

\bibpunct{(}{)}{;}{a}{,}{,}

\def\thesection{\arabic{section}}
\numberwithin{equation}{section}
\numberwithin{table}{section}

\def\wh{\widehat}

\def\cov{\mbox{Cov}}

\def\diag{\mbox{diag}}

\renewcommand{\thefootnote}{\fnsymbol{footnote}}

\newcommand{\vc}{\textnormal{vec}}

\newcommand{\bA}{{\mathbf A}}
\newcommand{\bB}{{\mathbf B}}
\newcommand{\bF}{{\mathbf F}}
\newcommand{\bE}{{\mathbf E}}
\newcommand{\bG}{{\mathbf G}}
\newcommand{\bH}{{\mathbf H}}
\newcommand{\bI}{{\mathbf I}}

\newcommand{\bQ}{{\mathbf Q}}

\newcommand{\bR}{{\mathbf R}}
\newcommand{\bS}{{\mathbf S}}

\newcommand{\bU}{{\mathbf U}}

\newcommand{\bW}{{\mathbf W}}
\newcommand{\bX}{{\mathbf X}}
\newcommand{\bY}{{\mathbf Y}}
\newcommand{\bZ}{{\mathbf Z}}
\newcommand{\ba}{{\mathbf a}}
\newcommand{\bb}{{\mathbf b}}

\newcommand{\bi}{{\mathbf i}}

\newcommand{\bu}{{\mathbf u}}
\newcommand{\bv}{{\mathbf v}}
\newcommand{\bw}{{\mathbf w}}

\newcommand{\by}{{\mathbf y}}

\newcommand{\balpha} {\boldsymbol{\alpha}}
\newcommand{\bbeta}  {\boldsymbol{\beta}}

\newcommand{\bgamma}{\boldsymbol{\gamma}}

\newcommand{\ve}{{\varepsilon}}
\renewcommand{\epsilon}{{\ve}}
\renewcommand{\hat}{\widehat}

\newcommand{\tr}{\mbox{tr}}

\def\JRSSB{{\sl Journal of the Royal Statistical Society}, {\bf B}}

\def\BKA{{\sl Biometrika}}
\def\JASA{{\sl Journal of the American Statistical Association}}

\newcommand{\astfootnote}[1]{%
\let\oldthefootnote=\thefootnote%
\setcounter{footnote}{0}%
\renewcommand{\thefootnote}{\fnsymbol{footnote}}%
\footnote{#1}%
\let\thefootnote=\oldthefootnote%
}

\begin{document}


\renewcommand{\baselinestretch}{2}

\markright{ \hbox{\footnotesize\rm 
}\hfill\\[-13pt]
\hbox{\footnotesize\rm
}\hfill }

\markboth{\hfill{\footnotesize\rm Hangjin Jiang, Baining Shen, Yuzhou Li and Zhaoxing Gao} \hfill}
{\hfill {\footnotesize\rm Regularized Estimation of MAR Models } \hfill}

\renewcommand{\thefootnote}{}
$\ $\par


\fontsize{12}{14pt plus.8pt minus .6pt}\selectfont \vspace{0.6pc}
\centerline{\large\bf Regularized Estimation of High-Dimensional Matrix-Variate Autoregressive Models}
\vspace{.1cm} \centerline{Hangjin Jiang$^1$, Baining Shen$^1$, Yuzhou Li$^1$, and  Zhaoxing Gao$^2$\astfootnote{Corresponding author: zhaoxing.gao@uestc.edu.cn (Z. Gao),  School of Mathematical Sciences, University of Electronic Science and Technology of China, Chengdu, 611731 P.R. China.}} \vspace{.1cm}
\centerline{\it $^1$Center for Data Science, Zhejiang University}
\centerline{\it
$^2$School of Mathematical Sciences, University of Electronic Science and Technology of China}
\vspace{.2cm} \fontsize{9}{11.5pt plus.8pt minus
.6pt}\selectfont

\vspace{-0.6cm}

\begin{quotation}
\noindent {\it Abstract:}
Matrix-variate time series data are increasingly popular in economics, statistics, and environmental studies, among other fields. This paper develops regularized estimation methods for analyzing high-dimensional matrix-variate time series using bilinear matrix-variate autoregressive models. The bilinear autoregressive structure is widely used for matrix-variate time series, as it reduces model complexity while capturing interactions between rows and columns. However, when dealing with large dimensions, the commonly used iterated least-squares method results in numerous estimated parameters, making interpretation difficult. To address this, we propose two regularized estimation methods to further reduce model dimensionality. The first assumes banded autoregressive coefficient matrices, where each data point interacts only with nearby points. A two-step estimation method is used: first, traditional iterated least-squares is applied for initial estimates, followed by a banded iterated least-squares approach. A Bayesian Information Criterion (BIC) is introduced to estimate the bandwidth of the coefficient matrices. The second method assumes sparse autoregressive matrices, applying the LASSO technique for regularization. We derive asymptotic properties for both methods as the dimensions diverge and the sample size 
$T\rightarrow\infty$. Simulations and real data examples demonstrate the effectiveness of our methods, comparing their forecasting performance against common autoregressive models in the literature.
\vspace{6pt}

\noindent {\it Key words and phrases:}
Matrix Time Series, High-dimension, Iterated Least-Squares, Band, Lasso
\par
\end{quotation}\par

\def\thefigure{\arabic{figure}}
\def\thetable{\arabic{table}}

\renewcommand{\theequation}{\thesection.\arabic{equation}}

\fontsize{12}{14pt plus.8pt minus .6pt}\selectfont

\setcounter{section}{0} 
\setcounter{equation}{0} 

\section{Introduction}

In recent years, with the development of advanced information technologies, modern data collection and storage capabilities have led to massive amounts of time series data. Multiple and high-dimensional time series are routinely observed in a wide range of applications, including economics, finance, engineering, environmental sciences, medical research, and others. In the past decades, various multivariate time series modeling methods have been studied in the literature. See \cite{Tsay_2014} and the references therein for details.
Recently, large tensor (or multi-dimensional array) time series data have become increasingly popular in the literature across various fields, including those mentioned. For example, a group of countries will report a set of economic indicators each quarter, forming a matrix-variate (2-dimensional array) time series, with each column representing a country and each row representing an economic indicator. To analyze large and high-dimensional datasets, dimension-reduction techniques have gained popularity for achieving efficient and effective analysis of high-dimensional time series data. Examples include the canonical correlation analysis (CCA) of \cite{BoxTiao_1977} and \cite{gaotsay2018a}, principal component analysis (PCA) of \cite{StockWatson_2002}, the scalar component model of \cite{TiaoTsay_1989}, and the factor model approach in \cite{BaiNg_Econometrica_2002}, \cite{StockWatson_2005}, \cite{forni2000,forni2005}, \cite{panyao2008}, \cite{LamYaoBathia_Biometrika_2011}, \cite{lamyao2012}, and \cite{gaotsay2020a,gaotsay2018b,gaotsay2021}, among others.
However, all the techniques developed for vector time series cannot be directly applied to matrix-variate time series, and simple vectorization of the matrix data often results in a significant number of estimated parameters, losing the original data structure. Therefore, further analysis methods should be developed to model such complex and dynamic datasets.

Recently, several methods have been developed for analyzing matrix-variate time series data, including factor models in \cite{wang2018}, \cite{chentsaychen2018}, \cite{yu2022}, and \cite{gaotsay2021c}, as well as the bilinear matrix-variate autoregressive model in \cite{chenxiaoyang2020}. To the best of our knowledge, only the method in \cite{chenxiaoyang2020} can be directly applied for out-of-sample forecasting, while others primarily focus on dimension reduction of the matrix data structures. Although \cite{chenxiaoyang2020} introduced effective techniques for estimating autoregressive coefficient matrices and explored their asymptotic properties, these methods are applicable only to matrix-variate time series data with fixed and small dimensions. Given that large-dimensional matrix-variate data are increasingly common in applications, the traditional iterated least-squares methods presented in \cite{chenxiaoyang2020} may not perform well, and the theoretical results may no longer hold. Therefore, new estimation methods must be considered in such contexts.

This paper represents an extension of the bilinear matrix-variate autoregressive model developed in \cite{chenxiaoyang2020} and the spatio-temporal data framework in \cite{hsu2021}. We focus on the scenario where the dimensions of matrix-variate data are growing, thereby extending the approach in \cite{chenxiaoyang2020} to high-dimensional contexts. To facilitate meaningful dimension reduction, we recognize that each observed data point interacts only with a limited number of others. For instance, spatio-temporal data points, such as PM$_{2.5}$ observations, may rely primarily on a few neighboring locations. More generally, each observation may dynamically depend on only a subset of other components. Our goal is to identify sparse autoregressive matrices that allow for further dimensional reduction while maintaining interpretability.

In this paper, we propose two regularized estimation methods to reduce the model's dimensions further. The first method assumes that the autoregressive coefficient matrices are banded, indicating that each observed data point interacts only with a limited number of neighboring points. We introduce a two-step estimation approach: the first step utilizes traditional iterated least-squares to obtain initial estimates, while the second step employs a banded iterated least-squares method. Additionally, we propose using the Bayesian Information Criterion (BIC) to estimate the bandwidths of the coefficient matrices.
The second method is similar but assumes that the autoregressive matrices are sparse, applying the LASSO technique for estimation. We derive the asymptotic properties of the proposed methods for diverging dimensions of the matrix-variate data as the sample size 
$T\rightarrow\infty$. Both simulated and real examples are used to evaluate the performance of our methods in finite samples, comparing them with commonly used techniques in the literature regarding the forecasting ability of autoregressive models.

This paper presents multiple contributions. First, the methods introduced in \cite{chenxiaoyang2020} are applicable only to matrix-variate time series data with fixed and relatively small dimensions. We extend this model to a high-dimensional environment, offering a broader perspective on matrix-autoregressive models that is increasingly relevant for practitioners as such data become more common in applications.
Second, coefficients obtained from traditional least-squares methods can be challenging to interpret due to the large number of parameters associated with higher dimensions. Our approaches, utilizing banded and general sparse structures, address this issue by facilitating meaningful dimensional reductions. The banded approach is particularly well-suited for analyzing spatio-temporal data, as the matrix structure corresponds to the locations of observations, making it reasonable to assume that each data point depends dynamically on only a few neighboring points.
Finally, we provide rigorous theoretical analysis, deriving the asymptotic properties of our proposed methods under these circumstances, thereby contributing to the theoretical foundation of this field.


The rest of the paper is organized as follows. We introduce the model and 
proposed estimation methodology in Section \ref{sec2} and 
study the theoretical properties of the proposed model and its associated 
estimates in Section \ref{sec3}. 
Numerical studies with both simulated and real data sets are given in Section \ref{sec4}, and 
Section \ref{sec5} provides some concluding remarks. 
All technical proofs are given in an online Supplementary Material. Throughout the article,
we use the following notation. For a $p\times 1$ vector
$\bu=(u_1,..., u_p)',$  $||\bu||_2 =\|\bu'\|_2= (\sum_{i=1}^{p} u_i^2)^{1/2} $
is the Euclidean norm, $\|\bu\|_\infty=\max_{1\leq i\leq p}|u_i|$ is the $\ell_\infty$-norm, and $\bI_p$ denotes a $p\times p$ identity matrix. For a matrix $\bH=(h_{ij})$, $\|\bH\|_1=\max_j\sum_i|h_{ij}|$, $\|\bH\|_\infty=\max_{i,j}|h_{ij}|$,  $\|\bH
\|_F=\sqrt{\sum_{i,j}h_{ij}^2}$ is the Frobenius norm, $\|\bH
\|_2=\sqrt{\lambda_{\max} (\bH' \bH ) }$ is the operator norm, where
$\lambda_{\max} (\cdot) $ denotes for the largest eigenvalue of a matrix, and $\|\bH\|_{\min}$ is the square root of the minimum non-zero eigenvalue of $\bH'\bH$. The superscript ${'}$ denotes 
the transpose of a vector or matrix. We also use the notation $a\asymp b$ to denote $a=O(b)$ and $b=O(a)$. $\mathbb{E}X$ denotes the expectation of random variable $X$, and $\bi_k$ be the unit vector with the $k$-th element equal to 1. Finally, $C$ is a constant having different values in different contexts.

\section{Model and Methodology}\label{sec2}
\subsection{Setting}
Let $\bY_t\in \mathbb{R}^{p_1\times p_2}$ be an observable $p_1\times p_2$ matrix-variate time series, we consider the matrix-variate autoregressive model of order $d\geq 1$ (MAR($d$)) introduced by \cite{chenxiaoyang2020} as follows:
\begin{equation}\label{md}
	\bY_t=\bA_1\bY_{t-1}\bB_1'+...+\bA_d\bY_{t-d}\bB_d'+\bE_t,
\end{equation}
where $\bA_i$ and $\bB_i$ are the coefficient matrices, and $\bE_t$ is a white noise term.  By a similar argument as that in traditional AR models, we may let $\bA=[\bA_1,...,\bA_d]$, $\bB=[\bB_1,...,\bB_d]$, and $\bG_t=\diag(\bY_{t-1},...,\bY_{t-d})$, the regression part in Model (\ref{md}) can be written as $\bA\bG_t\bB'$. Thus, 
without loss of generality, we only consider the case when $d=1$, i.e., study the following MAR(1) model:
\begin{equation}\label{md1}
	\bY_t=\bA\bY_{t-1}\bB'+\bE_t, 
\end{equation}
where $\bY_t \in \mathbb{R}^{p_1\times p_2}$, $\bA \in \mathbb{R}^{p_1\times p_1}$, and $\bB \in \mathbb{R}^{p_2\times p_2}$ are the coefficient matrices, and $\bE_t \in  \mathbb{R}^{p_1\times p_2}$ is the white noise term.

As discussed in \cite{chenxiaoyang2020}, the coefficient matrices $\bA$ and $\bB$ are not uniquely defined due to the identification issue. For example, $(\bA,\bB)$ can be replaced by $(c\bA,\bB/c)$ for some constant $c\neq 0$ without altering the equation in (\ref{md1}). Therefore, some identification conditions are required to impose on the coefficients. There are several ways to achieve this, for instance, we may assume $\|\bA\|_F=1$ and $\text{sign}(\tr(\bA))=1$ as that in \cite{hsu2021}.

\cite{chenxiaoyang2020} proposed  three methods to estimate the coefficient matrices when the dimensions $p_1$ and $p_2$ are fixed, which are (1) the Projection method, (2) Iterated least squares approach, and (3) Maximum likelihood estimation (MLE). Asymptotic properties of the estimators are also established therein. However, both their methods and asymptotic theory are derived under finite and fixed dimensions.

In this paper, we consider the estimation of the coefficient matrices in high-dimensional scenarios, i.e., $p_1,p_2\rightarrow\infty$ as $T\rightarrow\infty$. It is widely known that traditional methods usually fail when the dimensions are growing as the sample size increases, and one of the main reasons is that there will be much more parameters to be estimated. Therefore, some interpretable structures are often imposed in a high-dimensional framework.
Here, we present the estimation methods under two different cases: 1) the coefficient matrices $\bA$ and $\bB$ are banded ones,  and 2) $\bA$ and $\bB$ are sparse,  where the banded coefficients matrices are often used in spatio-temporal data when the observed value of one location only depends on those of a few neighborhoods. On the other hand, 
in the second scenario, we assume only a small proportion of elements in $\bA$ and $\bB$ are non-zero, which serves as a general sparsity condition. Our goal is to estimate the coefficients $\bA$ and $\bB$ under such conditions in a high-dimensional framework. We will discuss these two scenarios in the following sections.

\subsection{Estimation with the Banded Case}\label{sec22}
In this section, we consider the scenario that the coefficient matrices $\bA$ and $\bB$ are banded ones with bandwidths $k_1>0$ and $k_2>0$, respectively. That is,  we assume that
\begin{equation}\label{bands}
	a_{i,j}=0, b_{k,l}=0\,\,\text{for all}\,\, |i-j|>k_1,|k-l|>k_2,
\end{equation}
where $\bA=(a_{i,j})_{i,j=1}^{p_1}$, $\bB=(b_{k,l})_{k,l=1}^{p_2}$, and $k_1$ and $k_2$ are unknown bandwidths. Our goal is to estimate the coefficient matrices $\bA$ and $\bB$, and their corresponding bandwidth parameters $k_1$ and $k_2$.

We first assume that $k_1$ and $k_2$ are known, and we will propose a BIC approach to consistently estimate them in subsection \ref{sec:k} below. For the estimation of the coefficient matrices with banded structures, the procedure can be carried out in a similar way as the iterated least-squares method in \cite{chenxiaoyang2020}. 
Specifically, we first obtain initial estimators of $\bA$ and $\bB$ by the iterated least squares method proposed in \cite{chenxiaoyang2020}. Then we start with the initial estimators, and perform another iterated least squares method to estimate the banded coefficient matrices. For example, we may estimate $\bA$ and its bandwidths $k_1$ when the latest estimator for $\bB$ is given, and then estimate $\bB$ and its bandwidths $k_2$ by fixing the latest estimator for $\bA$. We repeat this procedure and the algorithm stops when the estimators converge. A description of the algorithm is outlined in \textbf{Algorithm 1}. Details on Steps 2(a) and 2(b) in \textbf{Algorithm 1} are given in the following subsections. 

\begin{algorithm}\label{alg1}
	\caption{Estimating algorithm for banded case}
	\begin{enumerate}
		\item  We use the iterated least-squares method in \cite{chenxiaoyang2020} to obtain the estimators $\wh\bA_0$ and $\wh\bB_0$ for $\bA$ and $\bB$, respectively. Denote the initial estimators as $\wh\bB^{(0)}=\wh\bB_0$ and $\wh\bA^{(0)}=\wh\bA_0$.
		\item For the $i$-th iteration ($i=1, 2, \cdots $),
		\begin{enumerate}
			\item  Fix the estimator $\wh\bB^{(i-1)}$ of $\bB$, the estimator $\wh\bA^{(i)}$ of $\bA$ is obtained by applying the least-squares method to Model (\ref{md1}). The estimator of the unknown bandwidth $k_1$, denoted by $\wh k_1^{(i)}$, is obtained based on a BIC given in section \ref{sec:k} below.
			\item Fix the estimator $\hat\bA^{(i)}$ of $\bA$, we estimate $\wh\bB^{(i)}$ and $\wh k_2^{(i)}$ using the same procedure as that in (a).
			\item The iteration stops if the convergence criterion is satisfied, otherwise we go to the next iteration and repeat Steps 2(a)--2(b).
		\end{enumerate}
	\end{enumerate}
\end{algorithm}
Note that either $\wh\bB^{(i-1)}$ in Step 2(a) or $\wh\bA^{(i)}$ in Step 2(b) needs to be normalized according to the identification conditions mentioned in Section 2.1. The initial estimator can be either $\wh\bB^{(0)}$ or $\wh\bA^{(0)}$ which does not influence the properties of the final estimators. For the convergence conditions, there are several useful ones that we can adopt in Step 2(c) of \textbf{Algorithm~1}. For example, we may take the following two convergence criteria:
\[\|\wh\bA^{(i)}-\wh\bA^{(i-1)}\|_F\leq \eta\,\,\text{and}\,\,\|\wh\bB^{(i)}-\wh\bB^{(i-1)}\|_F\leq \eta,\]
or
\[\|\wh\bB^{(i)}\otimes\wh\bA^{(i)}-\wh\bB^{(i-1)}\otimes\wh\bA^{(i-1)}\|_F\leq \eta,\]
where $\eta>0$ is a prescribed small constant. In practice, we may choose $\eta=10^{-6}$, and simulation results in Section \ref{sec4} suggest that our algorithm works well in finite samples.

\subsubsection{Iterated Least-Squares Estimation}\label{sec221}

Given $\bB=\wh\bB^{(i-1)}$, in order to obtain the estimator of $\bA$ and its bandwidth $k_1$, we write $\hat{\bQ}_t= \bY_t \hat{\bB}^{(i-1)}{'}$ and ${\bQ}_t= \bY_t \bB'$, and model (\ref{md1}) can be written as
\begin{equation}\label{md2a}
	\bY_t=\bA\hat{\bQ}_{t-1}+\bF_{1t}, \bF_{1t}=\bA({\bQ}_{t-1}-\hat{\bQ}_{t-1})+\bE_t.
\end{equation}
Let $\bA'=[\ba_1, \cdots, \ba_{p_1}]$, we have, 
$\bY_{t}'\bi_j=\hat{\bQ}_{t-1}'\ba_j+\bF_{1t}'\bi_j, j=1, 2, \cdots, p_1.$

Assuming the bandwidth of $\bA$ is $k$, then there are $\tau_j(k)$ non-zero elements in its $j$-th row $\ba_j$ of $\bA$, where
$$
\tau_j(k) = \left\{ \begin{array}{ll}
	k+j & {\rm if}~~~j  \leq k+1, \\
	2k+1 & {\rm if}~~~k+1 <j \leq p_1-k, \\
	p_1+k-j+1 & {\rm if}~~~p_1-k <j \leq p_1. \end{array}
\right.
$$

Let $\boldsymbol{\beta}_{j,k}$ be the $\tau_j(k) \times 1$ vector obtained by stacking non-zero elements in $\ba_j$, and $\bX_{j, t-1}^k$ be the corresponding $\tau_j(k)$ columns of $\hat{\bQ}_{t-1}'$. Denote $\bv_j=[\bi_j'\bY_2,\cdots, \bi_j'\bY_{T+1}]'\in \mathbb{R}^{Tp_2\times 1}$, $\bX_{j,k}=[\bX_{j, 1}^k, \cdots, \bX_{j, T}^k] \in \mathbb{R}^{Tp_2\times \tau_j(k)}$, and $\boldsymbol{f}_j=[\bi_j'\bF_{12},\cdots, \bi_j'\bF_{1,T+1}]'$, 
we have 
\begin{equation}\label{lin2a}
	\bv_j=\bX_{j,k} \boldsymbol{\beta}_{j,k}+ \boldsymbol{f}_j.
\end{equation}
Now, it follows from (\ref{lin2a}) that the least-squares estimator of $\boldsymbol{\beta}_{j,k}$ is denoted by $\hat{\boldsymbol{\beta}}_{j,k}=(\bX_{j,k}'\bX_{j,k})^{-1}\bX_{j,k}'\bv_j$, and the corresponding residual sum of squares can be written as
\begin{equation}\label{rss1}
	\text{RSS}_j(k, \ba_j)= \bv_j'(\bI - \bH_{\bX_{j,k}})\bv_j,
\end{equation}
where $\bH_{\bX_{j,k}}=\bX_{j,k}(\bX_{j,k}'\bX_{j,k})^{-1}\bX_{j,k}'$ is a hat matrix, which is a function of the unknown bandwidth $k$. 

Next, we consider the estimation of $\bB$ given $\hat{\bA}^{(i)}$. Similar to the technique used in (\ref{md2a}), let $\hat{\bR}_t= \hat{\bA}^{(i)}\bY_t $ and $\bR_t=\bA\bY_t $,
model (\ref{md1}) can be written as
\begin{equation}\label{md3}
	\bY_t=\hat{\bR}_{t-1}\bB'+\bF_{2t}, \bF_{2t}=(\bR_{t-1}- \hat{\bR}_{t-1})\bB'+\bE_t.
\end{equation}
Let $\bB'=(\bb_1, \cdots, \bb_{p_2})$, we have
$\bY_{t}\bi_j=\hat{\bR}_{t-1} \bb_j+\bF_{2t}\bi_j, j=1, 2, \cdots, p_2.$

Assuming the bandwidth of $\bB$ is $k$, then there are $\tau_j(k)$ non-zero elements in $\bb_j$, where
$$
\tau_j(k) = \left\{ \begin{array}{ll}
	k+j & {\rm if}~~~j  \leq k+1, \\
	2k+1 & {\rm if}~~~k+1 <j \leq p_2-k, \\
	p_2+k-j+1 & {\rm if}~~~p_2-k <j \leq p_2. \end{array}
\right.
$$
Let $\boldsymbol{\gamma}_{j,k}$ be the $\tau_j(k) \times 1$ vector obtained by stacking non-zero elements in $\bb_j$, and $\bG_{j, t-1}^k$ be the corresponding $\tau_j(k)$ columns of $\hat{\bR}_{t-1}$. Denote $\bw_j=[\bi_j'\bY_{2}',\cdots, \bi_j'\bY_{T+1}']'$, $\bG_{j,k}=[\bG_{j, 1}^k; \cdots; \bG_{j, T}^k]$, and $\boldsymbol{r}_j=[\bi_j'\bF_{22}',\cdots, \bi_j'\bF_{2,T+1}']'$, we have
\begin{equation}
	\bw_j=\bG_{j,k} \boldsymbol{\gamma}_{j,k}+ \boldsymbol{r}_j,
\end{equation}
and obtain the least-squares estimator of $\bgamma_{j,k}$ as $\hat{\boldsymbol{\gamma}}_{j,k}=(\bG_{j,k}'\bG_{j,k})^{-1}\bG_{j,k}'\bw_j$. The corresponding residual sum of squares is given by
\begin{equation}
	\text{RSS}_j(k,\bb_j)= \bw_j'(\bI - \bH_{\bG_{j,k}})\bw_j,     
\end{equation}
where $\bH_{\bG_{j,k}}=\bG_{j,k}(\bG_{j,k}'\bG_{j,k})^{-1}\bG_{j,k}'$ is a hat matrix, which is also a function of the unknown bandwidth $k$ as that in (\ref{rss1}).  

\subsubsection{Determining the bandwidth}\label{sec:k}
As discussed in Section \ref{sec221}, the estimation of the unknown coefficients depends on the bandwidth parameters $k_1$ and $k_2$, which are unknown in practice.
In this section,
we propose a Bayesian information criterion (BIC) to determine the unknown bandwidth of $\bA$. We first consider the estimation of the bandwidth $k_1$ of $\bA$.
For each prescribed $k_1\geq 1$, we may obtain the least-squares estimator of $\wh\bbeta_{j,k}$ from Model (\ref{lin2a}) as well as the residual-sum of squares in (\ref{rss1}). For $j=1,...,p_1$, we define 
\[\text{BIC}_j(k)=\log\text{RSS}_j(k,\ba_j)+\frac{C_{T_2}}{T_2}\tau_j(k) \log(p_1 \vee T_2), \]
where $T_2= p_2T$, and $C_{T_2}=\log \log (T_2)$. The bandwidth of the $j$-th row of $\bA$  estimated from $(\bX_{j,k}, \bv_j)$ is given by
\[ \hat{k}_{1,j}=\arg \min_{1 \leq k \leq K}\text{BIC}_j(k), \]
where $K$ is a prescribed upper bound of $k$ which may be taken as $\lceil T^{1/2} \rceil$. Finally, the estimated bandwidth of $\bA$ in the $i$-th iteration of {\bf Algorithm~1} is given by
$\hat{k}_1^{(i)}=\max_{1\leq j \leq p_1}  \hat{k}_{1,j} $, and the estimator for
$\bA$ in the $i$-th iteration is denoted by $\wh{\bA}^{(i)}=[\wh{\ba}_1,\wh{\ba}_2, \cdots, \wh{\ba}_{p_1}]'$, where the estimator $\wh{\ba}_j$ of $\ba_j$ is obtained by replacing the corresponding non-zero elements in $\ba_j$ by $\hat{\boldsymbol{\beta}}_{j,\hat{k}_j}$.

Similarly, for the estimation of the bandwidth parameter $k_2$, we can define the following BIC criterion
\[\text{BIC}_j(k)=\log\text{RSS}_j(k,\bb_j)+\frac{C_{T_1}}{T_1}\tau_j(k)\log(p_2 \vee T_1), \]
where $T_1= p_1T$, and $C_{T_1}=\log \log (T_1)$. The bandwidth of the $j$-th row of $\bB$ can be estimated by
\[ \hat{k}_{2,j}=\arg \min_{1 \leq k \leq K}\text{BIC}_j(k), \]
where $K$ is a prescribed upper bound of $k$ which may be taken as $\lceil T^{1/2} \rceil$. Finally, the bandwidth of $\bB$, is estimated as
$\hat{k}_2^{(i)}=\max_{1\leq j \leq p_2}  \hat{k}_{2,j} $ and $\bB$ is estimated as $\wh{\bB}^{(i)}=[ \wh{\bb}_1,\wh{\bb}_2, \cdots, \wh{\bb}_{p_2}]'$, where estimator $\wh{\bb}_j$ of $\bb_j$ is obtained by replacing corresponding non-zero elements in $\bb_j$ by $\hat{\boldsymbol{\beta}}_{j,\hat{k}_j}$, which is similar as that in estimating $k_1$.

In practice, the upper bound $K>0$ in the BICs defined above is a prescribed integer.
Our numerical results show that the procedure is insensitive
to the choice of $K$ so long as $K>k_1$ and $K>k_2$. In practice, we may take $K$ to be $\min([T^{1/2}],[p_1^{1/2}],[p_2^{1/2}])$  or choose $K$ by
checking the curvature of BIC$_i(k)$ directly.

\subsection{Estimation with Sparse Coefficient Matrices}\label{sec2.3}

In this section, we consider the estimation of the coefficient matrices in model~(\ref{md1}) under the scenario that the coefficients are sparse, i.e., we assume that $\bA$ and $\bB$ are sparse in the sense that only a few elements within are nonzero. Note that we may apply the properties of the Kronecker product to model (\ref{md1}), and  rewrite the model in the following two ways:
\begin{equation}\label{lso1}
	\vc(\bY_t)=((\bB\bY_{t-1}')\otimes\bI_{p_1})\vc(\bA)+\vc(\bE_t),
\end{equation}
and
\begin{equation}\label{lso2}
	\vc(\bY_t')=((\bA\bY_{t-1})\otimes\bI_{p_2})\vc(\bB)+\vc(\bE_t'),
\end{equation}
where $\vc(\cdot)$ is the vectorization operator that stacks all the columns of a matrix into a vector in order.

Let $\by_t=\vc(\bY_t)$ and $\by_t^*=\vc(\bY_t')$, it is possible to estimate $\bA$ when $\bB$ is known in (\ref{lso1}) and to estiamte $\bB$ when $\bA$ is known in (\ref{lso2}). For any consistent estimators $\wh\bB$ and $\wh\bA$ for $\bB$ and $\bA$, respectively, we may define $\bZ_{t-1}=(\bB\bY_{t-1}')\otimes\bI_{p_1}$ and $\wh\bZ_{t-1}=(\wh{\bB}\bY_{t-1}')\otimes\bI_{p_1}$. Similarly, we may also define $\bZ_{t-1}^*=(\bA\bY_{t-1})\otimes\bI_{p_2}$ and $\wh\bZ_{t-1}^*=(\wh{\bA}\bY_{t-1})\otimes\bI_{p_2}$. In view of the spare structures of the coefficient matrices, we may adopt some penalized method such as the Lasso to obtain the estimators.

Specifically, similar to the approach in the banded case in Section \ref{sec22}, we first obtain the initial estimators of $\bA$ and $\bB$ by applying the alternative least squares method proposed by \cite{chenxiaoyang2020}. Starting with these initial estimators, we may apply the Lasso technique to estimate $\bA$ by replacing $\bB$ with its latest estimator, and then  obtain the Lasso estimate of $\bB$ by replacing $\bA$ with its latest estimator. For example, denote the estimate for $\bB$ as $\wh\bB^{(i-1)}$ in the $i$-th iteration, we solve the following optimization problem:
\begin{equation}\label{soA}
	\wh\balpha=\arg\min_{\balpha\in R^{p_1^2}}\left\{\frac{1}{T}\sum_{t=2}^T\|\by_t-\wh\bZ_{t-1}\balpha\|_2^2+\lambda_{1,T}\|\balpha\|_1\right\},
\end{equation}
where $\wh\bB$ is equal to $\wh\bB^{(i-1)}$ in $\wh\bZ_{t-1}$ and $\lambda_{1,T}>0$ is a tuning parameter. Then the estimator $\wh\bA^{(i)}$ is obtained by reverting the $\wh\balpha$ to a $p_1\times p_1$ matrix according to the way $\vc(\cdot)$ is performed. Similarly, $\wh\bbeta$ is obtained by solving the following optimization problem:
\begin{equation}\label{soB}
	\wh\bbeta=\arg\min_{\bbeta\in R^{p_2^2}}\left\{\frac{1}{T}\sum_{t=2}^T\|\by_t^*-\wh\bZ_{t-1}^*\bbeta\|_2^2+\lambda_{2,T}\|\bbeta\|_1\right\},
\end{equation}
where $\wh\bA$ is equal to $\wh\bA^{(i)}$ in $\wh\bZ_{t-1}^*$. Then, the estimator $\wh\bB^{(i)}$ is obtained by reverting the $\wh\bbeta$ to a $p_2\times p_2$ matrix as before.
We can repeat this procedure until convergence. The estimation procedure is summarized in \textbf{Algorithm 2}. The convergence criteria are similar to those in {\bf Algorithm  1}, and we do not repeat them to save space.

\begin{algorithm}
	\caption{Estimating algorithm for sparse case}
	\begin{enumerate}
		\item Obtain $\wh\bA_0$ and $\wh\bB_0$ by the method in \cite{chenxiaoyang2020}, denoted as $\wh\bB^{(0)}=\wh\bB_0$ and $\wh\bA^{(0)}=\wh\bA_0$, respectively.
		\item For the $i$-th iteration ($i=1,2,...$),
		\begin{enumerate}
			\item Fix $\bB=\wh\bB^{(i-1)}$, apply Lasso to (\ref{soA}) and obtain $\wh\bA^{(i)}$,
			\item Fix $\bA=\wh\bA^{(i)}$, apply Lasso to (\ref{soB}) and obtain $\wh\bB^{(i)}$,
			\item The iteration stops if the convergence criterion is satisfied, otherwise we go to the next iteration and repeat Steps 2(a)--2(b).
		\end{enumerate}
	\end{enumerate}
\end{algorithm}

\section{Theoretical Properties}\label{sec3}
In this section, we establish the asymptotic properties of the estimators proposed in Section \ref{sec2}. We begin by outlining the regular conditions necessary for the theoretical proofs, followed by the statement of the asymptotic theorems. All proofs for the theorems are provided in an online Supplementary Material.

\subsection{Regular Conditions}
We introduce some notations first. 
A process $\vc(\bY_t)$ is $\alpha$-mixing if \[\alpha_p(k)=\sup_i \sup_{A \in \mathcal{F}_{-\infty}^i, B \in \mathcal{F}_{i+k}^{\infty}}|\mathbb{P}(A\cap B)-\mathbb{P}(A)\mathbb{P}(B)| \rightarrow 0,\]
where $\mathcal{F}_{l}^k$ is the $\sigma$-field generated by $\{\vc(\bY_t): l \leq t \leq k\}$. For $i=1,2$, 
define
$C_{i,\alpha}(S_i)=\{\Delta \in R^{p_i^2}: \|\Delta_{S_i^c}\|_1 \leq \alpha \|\Delta_{S_i}\|_1 \}$, where $S_i$ is a subset of $\{1,2,...,p_i^2\}$ and $\Delta_{S_i}$ is the vector of $\Delta$ restricted on the positions of $S_i$ and the other elements on indexes of $S_i^c$ are zero. Now, we introduce some assumptions for Model (\ref{md1}). 
\begin{itemize}
	\item[A1.] For ${\bY_t}=\{y_{ijt}\}$, we assume
	\begin{enumerate}
		\item[a.] The process ${\text{vec}(\bY_t)}$ is $\alpha$-mixing with the mixing coefficient satisfying the condition $\alpha_p(k) \leq \exp(-ck^{\gamma_1})$ for some $\gamma_1>0$. 
		\item[b.] $\sup_{i,j,t} \mathbb{P}(|y_{ij,t}|>s)\leq \exp(1-s^{\gamma_2})$ for $s>0$ and some  $\gamma_2>0$.
	\end{enumerate}
	\item[A2.] The innovations $\{\bE_t=(e_{ijt})\}$ are independent and identically distributed (i.i.d.) with mean 0, and
	\begin{enumerate}
		\item[a.] $\frac{1}{p_1}\sum_{i=1}^{p_1}{\mathbb{E}e_{ijt}^2} \rightarrow \sigma_1^2$ for $j=1,2 ,\cdots, p_2$ and $\frac{1}{p_2}\sum_{j=1}^{p_2}{\mathbb{E}e_{ijt}^2} \rightarrow \sigma_2^2$, for $i=1, 2, \cdots, p_1$.
		\item[b.] $\sup_{i,j,t} \mathbb{P}(|e_{ij,t}|>s)\leq \exp(1-s^{\gamma_3})$ for $s>0$ and some  $\gamma_3>0$.
	\end{enumerate}
	\item[A3.] $\rho(\bA)\rho(\bB)<1$, where $\rho(\bA)$ and $\rho(\bB)$ are the spectral radii of $\bA$ and $\bB$, respectively.
	\item [A4.]  For any two sub-columns of $\bQ_t'$ (\text{or} $\bR_t$), denoted by $\bW_t$ and $\bU_t$, and $\bW_t \neq \bU_t$, let $p=p_2$ (\text{or} $p_1$),  $\Sigma_{\bU}=p^{-1}\mathbb{E}\bU_t'\bU_t$, $\Sigma_{\bW\bU}=p^{-1}\mathbb{E}\bW_t'\bU_t$, and $\Sigma_{\bW}=p^{-1}\mathbb{E}\bW_t'\bW_t$, there exists some positive constants $\lambda_1 \leq \lambda_2$, such that
	$\lambda_1 \leq \lambda_{\min}(\Sigma_{\bU})\leq \lambda_{\max}(\Sigma_{\bU}) \leq \lambda_2$, and $\lambda_1 \leq \lambda_{\min}(\Sigma_{\bW}- \Sigma_{\bW\bU}\Sigma_{\bU}^{-1}\Sigma_{\bW\bU}') \leq \lambda_{\max}(\Sigma_{\bW}- \Sigma_{\bW\bU}\Sigma_{\bU}^{-1}\Sigma_{\bW\bU}') \leq \lambda_2$.
	\item[A5.] For the banded matrix $\bA$, $|a_{i, i-k_1}|$ or $|a_{i, i+k_1}|$, $i=1, 2, \cdots, p_1$, is greater than $\{C_{T_2}k_1T_2^{-1}\log(p_1 \vee T_2)\}^{1/2}$ with $T_2=p_2T$ and $C_{T_2}=\log \log T_2$; Similarly, for the banded matrix $\bB$, $|b_{i, i-k_2}|$ or $|b_{i, i+k_2}|$, $i=1, 2, \cdots, p_2$, is greater than $\{C_{T_1}k_2T_1^{-1}\log(p_2 \vee T_1)\}^{1/2}$ with $T_1=p_1T$ and $C_{T_1}=\log \log T_1$.
	\item[A6.] For matrices $\bA$ and $\bB$, $a_{i,j}$ and $b_{i,j}$ are bounded uniformly, and  $\|\bA^k\|_2 \leq \delta^k$ and $\|\bB^k\|_2 \leq \delta^k$ for $k \geq 2$, where $\delta \in (0,1)$ is independent of $p_1$, $p_2$, and $k$.
	\item[A7.] Let $S_0$ be a subset of $\{1,2,...,p_1^2\}$ with cardinality $s_0$ consisting of the indexes of the non-zero components in $\boldsymbol{\alpha}=\vc(\bA)$, and $S_0^c$ be its complement. Let $\bZ_t=(\bB\bY_t')\otimes \bI_{p_1}$, (a) when $p_1$ is finite, there exists a constant $C_2 > 0$ such that $\lambda_{min}\{\mathbb{E}(\bZ_t\bZ_t')\} > C_2$; (b) when $p_1$ is diverging, the matrix $\bZ := (\bZ_1,...,\bZ_T)'$ satisfies the restricted eigenvalues condition, $\frac{1}{T}\|\bZ\Delta\|_2^2 \geq \kappa \|\Delta\|_2^2$, for all $\Delta \in C_{1,3}(S_0)$. Similar assumptions also hold for $\bZ_t^*$ defined in Section 2.3.
	\item[A8.] For matrices $\bA$ and $\bB$, $\lambda_1< \lambda_{\min}\{p_2^{-1}\mathbb{E}[(\bA\bY_t)\otimes(\bY_t\bB')]\}<\lambda_{\max}\{p_2^{-1}\mathbb{E}[(\bA\bY_t)\otimes(\bY_t\bB')]\} <\lambda_2$.
\end{itemize}

Conditions A1(a-b) are standard for econometric time series models. Condition A2(a) ensures the row and column variances of $\bE_t$ exits, and  condition A2(b) is used to bound a new time series built on $\bE_t$ and $\bY_t$.
Condition A3 ensures that model (\ref{md1}) is stationary and causal, as shown in proposition 1 in \cite{chenxiaoyang2020}.  
Conditions A4-A7 are imposed to prove the consistency of the estimated bandwidth by BIC in Section \ref{sec:k}. Condition A5 ensures that the bandwidth is asymptotically identifiable, since both $\{C_{T_2}k_1T_2^{-1}\log(p_1 \vee T_2)\}^{1/2}$ and $\{C_{T_1}k_2T_1^{-1}\log(p_2 \vee T_1)\}^{1/2}$ is the minimum magnitude of a non-zero coefficient to be identifiable, see, e.g. \cite{gaoetal2017}.
Condition A7(a) indicates that the regressors have a non-singular covariance and the least-squares estimators are well defined when $p_1$ is finite. Condition A5(b) is the well-known restricted-eigenvalue condition in Lasso regressions; see Chapter 6 in Buhlmann and Van De Geer (2011). The condition in Condition A7(b) can also be replaced by a more general Restricted Strong Convexity condition that is commonly used in high-dimensional regularized estimation problems. See Chapter 9 of Wainwright (2019) for details. 

\subsection{Asymptotic properties}
In this section, we study the theoretical properties of the proposed method, i.e., the convergence of \textbf{Algorithm~1} and \textbf{Algorithm~2} in Section 2. Since we take estimators, $\hat{\bA}_0$ and $\hat{\bB}_0$, from the iterated least squares in \cite{chenxiaoyang2020} as our initials in \textbf{Algorithm~1} and \textbf{Algorithm~2}, we first study the convergence rate of $\hat{\bA}_0$ and $\hat{\bB}_0$.  Let $\Sigma$ be the covariance matrix of $\vc(\bE_t)$, and define $\boldsymbol{H}=\mathbb{E}\left(\boldsymbol{W}_t \boldsymbol{W}_t^{\prime}\right)+\boldsymbol{\gamma} \boldsymbol{\gamma}^{\prime}$, where $\boldsymbol{W}_t^{\prime}=[(\bB\bY_t^{\prime})\otimes\bI:\bI\otimes(\bA\bY_t)]\in \mathbb{R}^{p_1p_2 \times (p_1^2+p_2^2)}$ and $\boldsymbol{\gamma}=(\vc(\bA)^{\prime}, \boldsymbol{0})'\in \mathbb{R}^{p_1^2+p_2^2}$.  Let $p=\max\{p_1, p_2\}$,
we have following result for $\hat{\bA}_0$ and $\hat{\bB}_0$. 

\begin{proposition}
	Let conditions A1-A3 hold. If $\bA$, $\bB$, $\Sigma$ are nonsingular, $\lambda_{\min}(\bH) \geq \lambda_h >0$, and $T, p_1, p_2 \rightarrow \infty$, 
	then, $$\|\hat{\bA}_0-\bA\|_F^2+\|\hat{\bB}_0-\bB\|_F^2= O_p(\frac{p_1^2p_2}{T}+\frac{p_2^2p_1}{T}).$$
\end{proposition}
\begin{remark} (1) When the dimensions of $\bY_t$, $p_1$ and $p_2$, are fixed, the convergence rate of $\|\hat{\bA}_0-\bA\|_F^2$ and $\|\hat{\bB}_0-\bB\|_F^2$ are both of order $O_p(1/T)$. This recovers the low-dimensional case.\\
	(2) According to this Proposition, we have $\|\hat{\bA}_0-\bA\|_F^2+\|\hat{\bB}_0-\bB\|_F^2 \rightarrow 0$ if $p_1^2p_2/T \rightarrow 0$ and $p_2^2p_1/T \rightarrow 0$, which can be simplified as $pp_1p_2/T \rightarrow 0$.
\end{remark}

Next, we show the convergence rate of estimators for $\bA$ and $\bB$ in \textbf{Algorithm~1} by assuming the bandwidth $k_1$ and $k_2$ are known and fixed.  Theorem \ref{tm1} below shows that estimators from {\bf Algorithm~1} are consistent when $p^3/T \rightarrow 0$.

\begin{theorem}\label{tm1} Assume conditions A1-A6 and A8 hold. Let $\hat{\bB}^{(i-1)}$ be the latest estimator of $\bB$ in \textbf{Algorithm 1} with $\hat{\bB}^{(0)}=\hat{\bB}_0$ and $i\geq 1$.  We have, for $i>1$,
	$$\|\hat{\bA}^{(i)} - \bA\|_F^2=O_p(\eta) \text{ and } \|\hat{\bB}^{(i)} - \bB\|_F^2=O_p(\eta),$$
	and,
	$$\|\hat{\bA}^{(i)} - \bA\|_2^2=O_p(\eta) \text{ and } \|\hat{\bB}^{(i)} - \bB\|_2^2=O_p(\eta),$$
	where $\eta=\|\hat{\bB}^{(i-1)}-\bB\|_F^2$.
\end{theorem}
\begin{remark} (i) In this theorem, \textbf{Algorithm~1} is assumed to begin estimating 
	$\bA$ by fixing 
	$\bB$ at its most recent estimate. However, the same conclusion will hold if we reverse the estimation order.\\
	(ii) In the first iteration, we take $\hat{\bB}^{(0)}=\hat{\bB}_0$, by Proposition 1, we have $\|\hat{\bB}_0-\bB\|_F^2 \rightarrow 0$ when $pp_1p_2/T \rightarrow 0$. The condition  $pp_1p_2/T \rightarrow 0$ also ensures that the error of $\bA^{(i)}$ is primarily influenced by the error arising from approximating 
	$\bB$ with its latest estimate.
\end{remark}

Theorem~\ref{tm1} is based on the assumption that the bandwidths are known, which is often not the case in real-world problems. However, if consistent estimators for these two unknown bandwidths exist, Theorem~\ref{tm1} remains valid. We will now demonstrate the consistency of the estimated bandwidths in \textbf{Algorithm~1}.
\begin{theorem}\label{tm2}
	Assume conditions A1-A6 hold. Let $\hat{\bB}^{(i-1)}$ be the latest estimator of $\bB$ in \textbf{Algorithm 1} with $\hat{\bB}^{(0)}=\hat{\bB}_0$. For $i>1$, if $p^5/T \rightarrow 0$, then 
	$$P(\hat{k}_1^{(i)}= k_1) \rightarrow 1, \text{ and } P(\hat{k}_2^{(i)}= k_2) \rightarrow 1, \text{ as } T, p_1, p_2 \rightarrow \infty.$$
\end{theorem}

\begin{remark} (i) In Theorem~\ref{tm1}, $k_1$ and $k_2$ are assumed to be fixed, since model (\ref{md1}) is useful only when $k_1$ and $k_2$ are small and finite. However, Theorem \ref{tm1} still holds when $k_1$ and $k_2$ diverges to $\infty$ along with $T, p_1, p_2$ so long as $k_1=o\{C_{T_2}^{-1}T_2/\log(p_1 \vee T_2)\}$, and $k_2=o\{C_{T_1}^{-1}T_1/\log(p_2 \vee T_1)\}$, where $T_2=p_2T$, $C_{T_2}=\log \log T_2$, $T_1=p_1T$, and $C_{T_1}=\log \log T_1$. See the proof of Theorem \ref{tm1} in Supplementary Material. \\
	(ii) In this theorem, \textbf{Algorithm~1} is assumed to begin estimating 
	$\bA$ by fixing 
	$\bB$
	at its most recent estimate. However, the same conclusion holds if we reverse the estimation order.

\end{remark}

Next, we examine the convergence rate of \textbf{Algorithm2} for estimating 
$\bA$ and 
$\bB$ in the sparse case. Similar to \textbf{Algorithm1}, \textbf{Algorithm~2} also begins with 
$\hat{\bA}_0$ and $\hat{\bB}_0$, which are derived from the iterated least squares method in \cite{chenxiaoyang2020}.

\begin{theorem}\label{tm3} Assume conditions A1-A3 and A6-A7 hold. Let $\hat{\bB}^{(i-1)}$ be the latest estimator of $\bB$ in \textbf{Algorithm 2} with $\hat{\bB}^{0}=\hat{\bB}_0$ and $i\geq 1$, we have,
	$$\|\hat{\bA}^{(i)} - \bA\|_2=O_p(\lambda_1\sqrt{s_0})\text{ and } \|\hat{\bB}^{(i)} - \bB\|_2 =O_p( \lambda_2\sqrt{s_0}),$$
	where $\lambda_1, \lambda_2 \asymp \sqrt{\frac{p_1p_2p}{T}}$, and $s_0=|S_0|$.
\end{theorem}
Similar to the results in the banded case, Theorem~\ref{tm3} shows that the estimated sparse coefficients are consistent under the condition that $pp_1p_2/T\rightarrow 0$ for finite sparsity $0<s_0< \infty$.
\section{Numerical Results}\label{sec4}
In this section, we examine the finite sample properties of the proposed method and provide real data examples to evaluate its forecasting performance compared to the alternative least-squares (ALSE) method proposed by \cite{chenxiaoyang2020}. In Section \ref{sec4.1}, we conduct Monte Carlo experiments to demonstrate the convergence of the proposed methods in estimating the coefficient matrices under two scenarios, alongside comparisons with the estimators obtained using the ALSE method. In Section \ref{sec4.2}, we apply our method to two real data examples.
\subsection{Simulation}\label{sec4.1}
This section outlines our methodology for evaluating the finite sample properties of the proposed methods through Monte-Carlo experiments. The observed data matrix $\bY_t$ are simulated from model~(\ref{md1}) under different conditions for matrices $\bA$ and $\bB$, and each entry in the white noise $\bE_t$ is generated from the standard normal distribution, with $\cov(\text{vec}(\bE_t)) = \bI_{p_1p_2}$. Through these simulations, we aim to demonstrate the convergence of our proposed methods in comparison with the ALSE method, as the sample size increases. Furthermore, we examine the accuracy of our proposed methods in estimating unknown bandwidths under the banded cases. The impact of penalty parameters on the estimation results is also investigated under the sparse cases. All of the results are obtained by conducting $100$ independent replications.

To study the convergence of the estimators for matrices $\bA$ and $\bB$, some identification conditions are required to impose on the coefficients. In this experiment, we assume $\|\bA\|_F = 1$ and $ \text{sign}(\text{tr}(\bA)) = 1$. The convergence criteria of the iterations in our algorithms are specified as 
$ \| \hat{\bA}^{(i+1)} - \hat{\bA}^{(i)} \| _F < 10^{-6} $ and
$\| \hat{\bB}^{(i+1)} - \hat{\bB}^{(i)} \| _F < 10^{-6}$.


\subsubsection{Banded case}\label{sec4.11}
This section presents a comprehensive analysis to investigate the performance of \textbf{Algorithm 1}. We will study the convergence of the estimators under the scenarios that we start with either $\wh\bA^{(0)}$ or $\wh\bB^{(0)}$ as initial estimates. Additionally, we examine the accuracy of the bandwidth estimation and the algorithm's convergence as the sample size 
$T$ increases.


For each configuration of $(p_1,p_2,k_1,k_2,T)$, we generate $\bY_t$ according to model~(\ref{md1}).
Specifically, for given dimensions $p_1$, $p_2$, and bandwidths $k_1$ and $k_2$, the observed data $\bY_t$ are simulated according to model (\ref{md1}), where the entries of $\bA$ and $\bB$ are generated as follows: (1) for entries of $\bA$, $\{a_{i,j}: |i-j| \le k_1 \} $ are generated independently from $ U[-1, 1]$, and other elements are zero. We re-scale $ \bA $ such that $ \| \bA \|_F$ = 1; (2) for entries of $\bB$, $\{b_{i,j}: |i-j| \le k_2 \} $ are generated independently from $ U[-1, 1] $, and other elements are zero. We re-scale $\bB$ so that $ \rho=\rho (\bA) \rho (\bB) = 0.5 $. The white noise $\bE_t$ are generated from standard normal distribution with $\cov(\text{vec}(\bE_t)) = \bI_{p_1p_2}$.

Firstly, we examine the convergence of \textbf{Algorithm 1} is insensitive to the choice of the initial estimators. Note that there are two iteration orders that may occur in {\bf Algorithm 1}. We may first estimate $\bA$ for given initial estimator $\wh\bB^{(0)}$, or estimate $\bB$ for given $\wh\bA^{(0)}$. We denote the estimated coefficient matrices via these two procedures by $ (\hat{\bA}_1,\hat{\bB}_1 ) $ and $ (\hat{\bA}_2,\hat{\bB}_2 ) $, respectively. 
The dimensions are set as $(p_1, p_2)=(6, 4) ,(8, 5)$ and $(9, 6) $ with bandwidths $(k_1, k_2) = (2, 1) $ for each $(p_1,p_2)$. The mean, median, and maximum of $\log_{10}{(\|\hat{\bA}_1 - \hat{\bA}_2\|_F ) }$ and $\log_{10}{(\|\hat{\bB}_1 - \hat{\bB}_2\|_F ) }$ are reported in Table \ref{tb1}. From Table~\ref{tb1} we see that the convergence of the estimators is insensitive to the choice of the initial estimators that we use in {\bf Algorithm 1}. On the other hand, the reported errors  in Table~\ref{tb1} are all less than $-6$, which is in accordance with the convergence criteria where the upper bound $\eta$ is chosen as $10^{-6}$ in Section \ref{sec2}. 
\begin{table}
	\caption{The mean, median, and maximum of $\log_{10}{(\|\hat{\bA}_1 - \hat{\bA}_2\|_F ) }$ and $\log_{10}{(\|\hat{\bB}_1 - \hat{\bB}_2\|_F ) }$, where $ (\hat{\bA}_1,\hat{\bB}_1 ) $ and $ (\hat{\bA}_2,\hat{\bB}_2 ) $ are obtained by starting with the initial estimators $\wh\bB^{(0)}$ and $\wh\bA^{(0)}$ in Algorithm 1, respectively.}
	\label{tb1}
	\centering
	\begin{tabular}{ccccccc}
		\toprule
		& \multicolumn{3}{c}{$\log_{10}{(\|\hat{\bA}_1 - \hat{\bA}_2\|_F ) }$} & \multicolumn{3}{c}{$\log_{10}{(\|\hat{\bB}_1 - \hat{\bB}_2\|_F ) }$} \\ 
		\cmidrule(lr){2-4} \cmidrule(lr){5-7}
		$(p_1, p_2)$ & Mean                & Median              & Maximum              & Mean                & Median              & Maximum              \\ \midrule
		$(6, 4)$     & -7.67               & -7.75               & -7.09                & -6.64               & -6.85               & -6.39                \\
		$(8, 5)$     & -7.57               & -7.62               & -7.19                & -6.38               & -6.43               & -6.01                \\
		$(9, 6)$     & -7.42               & -7.52               & -6.91                & -6.23               & -6.36               & -6.00      \\ \bottomrule         
	\end{tabular}
\end{table}

Second, we show the accuracy of \textbf{Algorithm 1} in estimating the unknown bandwidths $k_1$ and $k_2$.  The empirical frequencies of the events $\{\hat{k}_1 = k_1\} $ and $\{\hat{k}_2 = k_2\}$ are reported in Table~\ref{tb2}, where we set $(k_1,k_2)=(1,1)$ and $(2,1)$ and the sample size $T=100,200,400$ and $800$. For a fixed sample size $T$ and the bandwidths $(k_1,k_2)$, the dimensions $(p_1,p_2)$ are set to $(6,4)$, $(8,5)$, and $(9,6)$.
Results in Table \ref{tb2} show that the accuracy of estimated $\hat{k}_1$ and $\hat{k}_2$ is pretty satisfactory, and it increases with sample size $T$ for each $(p_1,p_2,k_1,k_2)$ in most cases. 

\begin{table}[htp]
	\caption{Accuracy of \textbf{Algorithm 1} in estimating unknown bandwidths under different settings, where $E_1$ and $E_2$ represent the empirical frequencies of the events $\{\hat{k}_1 = k_1\}$ and $\{\hat{k}_2 = k_2\}$, respectively. }
	\label{tb2}
	\centering
	\begin{tabular}{cccccccccc}
		\toprule
		& \multicolumn{2}{c}{$T = 100$} & \multicolumn{2}{c}{$T = 200$} & \multicolumn{2}{c}{$T=400$} & \multicolumn{2}{c}{$T = 800$} \\ 
		\cmidrule(lr){2-3} \cmidrule(lr){4-5} \cmidrule(lr){6-7} \cmidrule(lr){8-9} 
		$(p_1, p_2)$ & $E_1$         & $E_2$         & $E_1$         & $E_2$         & $E_1$        & $E_2$        & $E_1$         & $E_2$         \\ \midrule
		\multicolumn{9}{c}{$(k_1, k_2) = (1, 1)$}                                                                                                  \\
		$(6, 4)$     & 100           & 100           & 100           & 100           & 100          & 100          & 100           & 100           \\
		$(8, 5)$     & 100           & 99            & 100           & 99            & 100          & 100          & 100           & 100           \\
		$(9, 6)$     & 98            & 99            & 100           & 99            & 100          & 100          & 100           & 100           \\ \midrule
		\multicolumn{9}{c}{$(k_1, k_2) = (2, 1)$}                                                                                                  \\ 
		$(6, 4)$     & 98            & 100           & 100           & 100           & 100          & 100          & 100           & 100           \\
		$(8, 5)$     & 99            & 100           & 100           & 100           & 100          & 100          & 100           & 100           \\
		$(9, 6)$     & 99            & 100           & 100           & 100           & 100          & 100          & 100           & 100           \\  \bottomrule
	\end{tabular}
\end{table}
Next, we show the convergence pattern of \textbf{Algorithm 1} under different configurations of $(p_1,p_2,k_1,k_2)$ as the sample size $T$ increases. We also compare the estimation accuracy with the ALSE method in \cite{chenxiaoyang2020}. The estimation errors for $\bA$ and $\bB$, denoted by $ \log(\|\hat{\bA} - \bA \|_F) $ and $ \log(\|\hat{\bB} - \bB \|_F) $, respectively, are reported in Table~\ref{tb3}, where $(p_1,p_2)=(6,4)$ and $(9,6)$, the sample size $T=200,500,1000,2000$, and the bandwidths $(k_1,k_2)=(2,1)$. For each setting, we consider two scenarios that $\rho(\bA)\rho(\bB)=0.5$ and $0.8$ to show the results are consistent for different strengths of the coefficient matrices. From Table~\ref{tb3}, we see that estimation errors obtained by the proposed method and the ALSE all decrease as the sample size increase for each configuration, which is in line with our theoretical results. On the other hand, we also see that the estimation error obtained by our proposed method is smaller than that  by the ALSE, implying that our estimation procedure is more accurate than the ALSE.

\begin{figure}[htp]
	\centering
	\includegraphics[scale=0.5]{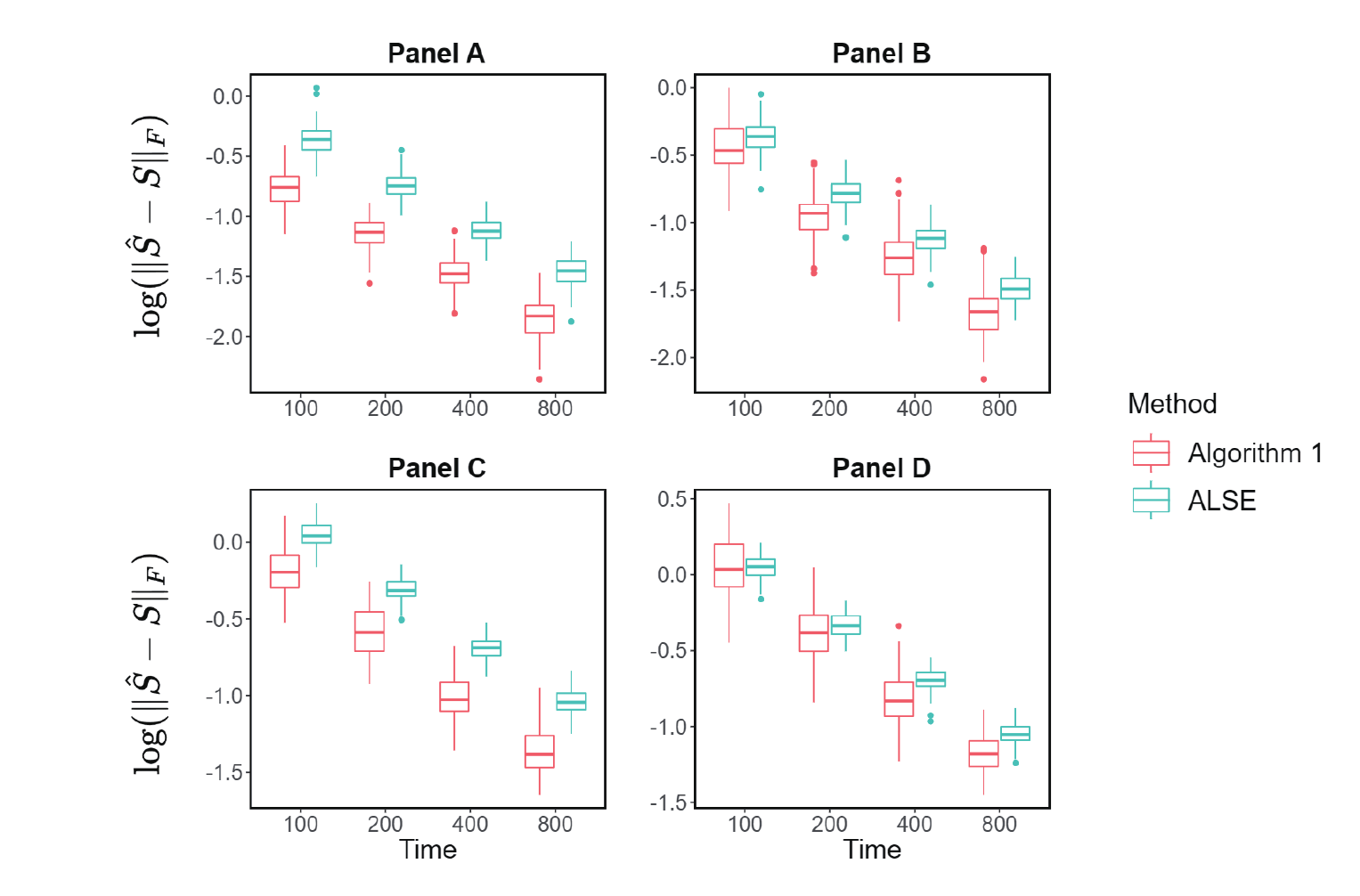}
	\caption{ Error of Kronecker product $\log(\|\hat{\bS} - \bS \|_F)$ of ALSE and \textbf{Algorithm 1} under different settings of ($p_1, p_2$) and $(k_1, k_2)$.  \textbf{Panel A}: $(p_1, p_2) = (6, 4)$, $(k_1, k_2) = (1, 1)$; \textbf{Panel B}:
		$(p_1, p_2) = (6, 4)$, $(k_1, k_2) = (2, 1)$; \textbf{Panel C}: $(p_1, p_2) = (9, 6)$, $(k_1, k_2) = (2, 1)$; \textbf{Panel D}: $(p_1, p_2) = (9, 6)$, $(k_1, k_2) = (3, 2)$; Results are reported from 100 independent repetitions.}
	\label{fig1}
\end{figure}

Furthermore, we define $ \bS := \bB \otimes \bA $ and $ \wh\bS := \wh\bB \otimes \wh\bA $, using the distance $ \log(|\wh\bS - \bS |_F) $ to evaluate the overall performance of our proposed method. For simplicity, we set the dimensions $(p_1, p_2) = (6, 4)$ and $(9, 6)$, and the bandwidths $(k_1, k_2)$ to $(1, 1)$ and $(2, 1)$ for each $(p_1, p_2)$. We fix $ \rho(\bA)\rho(\bB) = 0.5$ in this experiment, and the boxplots of the estimated errors $ \log(|\wh\bS - \bS |_F) $ are shown in Figure~\ref{fig1}.
It is clear that both methods converge under these settings, and our proposed \textbf{Algorithm 1} performs better than the ALSE method in terms of estimation errors, which aligns with our theoretical results.

\begin{table}[htp]\small
	\centering
	\caption{ The average estimation errors of the coefficient matrices by \textbf{Algorithm 1} and ALSE.}
	\label{tb3}
	\begin{tabular}{cccccccccc}
		\toprule
		&  & \multicolumn{4}{c}{Algorithm 1}    & \multicolumn{4}{c}{ALSE}           \\
		\cmidrule(lr){3-6} \cmidrule(lr){7-10}                  
		$(p_1, p_2)$  &  $\rho$  & T = 200 & 500    & 1000   & 2000   & T = 200 & 500    & 1000   & 2000\\ \midrule         
		\multicolumn{10}{c}{$ \log(\| \hat{\bA} - \bA \|_F) $}                                \\                      
		$(6, 4)$                      & \multicolumn{1}{c}{0.5}           & -1.960  & -2.454 & -2.842 & -3.214 & -1.830  & -2.298 & -2.637 & -2.999 \\
		$(6, 4)$                      & \multicolumn{1}{c}{0.8}           & -2.536  & -3.022 & -3.354 & -3.677 & -2.295  & -2.777 & -3.097 & -3.462 \\
		$(9, 6)$                      & \multicolumn{1}{c}{0.5}           & -1.191  & -1.691 & -2.236 & -2.733 & -1.215  & -1.666 & -2.009 & -2.348 \\
		$(9, 6)$                      & \multicolumn{1}{c}{0.8}           & -2.115  & -2.499 & -2.813 & -3.089 & -1.798  & -2.267 & -2.596 & -2.949 \\\midrule      
		\multicolumn{10}{c}{$ \log(\| \hat{\bB} - \bB \|_F) $}                                                                                         \\
		$(6, 4)$                      & \multicolumn{1}{c}{0.5}           & -1.701  & -2.154 & -2.474 & -2.805 & -1.401  & -1.882 & -2.208 & -2.558 \\
		$(6, 4)$                      & \multicolumn{1}{c}{0.8}           & -1.758  & -2.232 & -2.564 & -2.945 & -1.479  & -1.961 & -2.318 & -2.696 \\
		$(9, 6)$                      & \multicolumn{1}{c}{0.5}           & -1.314  & -1.799 & -2.154 & -2.513 & -0.860  & -1.328 & -1.704 & -2.058 \\
		$(9, 6)$                      & \multicolumn{1}{c}{0.8}           & -1.444  & -1.967 & -2.268 & -2.615 & -0.987  & -1.477 & -1.826 & -2.175 \\ \midrule      
		\multicolumn{10}{c}{$ \log(\| \hat{\bS} - \bS \|_F) $}                                                                                         \\
		$(6, 4)$                      & \multicolumn{1}{c}{0.5}           & -0.950  & -1.434 & -1.806 & -2.166 & -0.778  & -1.252 & -1.591 & -1.946 \\
		$(6, 4)$                      & \multicolumn{1}{c}{0.8}           & -1.139  & -1.617 & -1.953 & -2.295 & -0.892  & -1.369 & -1.705 & -2.076 \\
		$(9, 6)$                      & \multicolumn{1}{c}{0.5}           & -0.394  & -0.885 & -1.388 & -1.855 & -0.290  & -0.753 & -1.110 & -1.454 \\
		$(9, 6)$                      & \multicolumn{1}{c}{0.8}           & -0.735  & -1.151 & -1.464 & -1.751 & -0.381  & -0.858 & -1.195 & -1.546 \\ \bottomrule
	\end{tabular}
\end{table}

\subsubsection{Sparse case}

{This section evaluates the performance of \textbf{Algorithm 2} in estimating sparse coefficient matrices. First, we assess its ability to recover the non-zero elements of matrices $\bA$ and $\bB$. The tuning parameters $\lambda_{1,T}$ and $\lambda_{2,T}$ in equations (\ref{soA}) and (\ref{soB}) significantly influence the sparsity of $\bA$ and $\bB$, respectively. It is important to note that if these parameters are adjusted in each iteration of \textbf{Algorithm 2}, the algorithm will not converge, as the objective function changes with the tuning parameters. Therefore, we select the tuning parameters in the first iteration and keep them fixed for subsequent iterations.
	In practice, tuning parameters are usually chosen through cross-validation (CV). The R package glmnet offers two options: (1) sdCV, which selects $\lambda_{1,T}$ ($\lambda_{2,T}$) as the largest value of $\lambda$ such that the corresponding CV error is within 1 standard error of the minimum, and (2) mCV, which selects $\lambda_{1,T}$ ($\lambda_{2,T}$) as the value of $\lambda$ that minimizes the CV error.
	
	Alternatively, tuning parameters can also be selected based on variable selection stability, as discussed in \cite{stable2010} and \cite{Sun2013}. The key idea is to choose tuning parameters that ensure stability in the variable selection process of the penalized regression model. We employ the Kappa Selection Criterion (KSC) proposed by \cite{Sun2013} to select $\lambda_{1,T}$ and $\lambda_{2,T}$. Here, variable selection stability is defined as the expected value of Cohen’s kappa coefficient \citep{cohen1960} between active sets obtained from two independent and identical datasets.
	For instance, consider problem~(\ref{soA}). Given $\lambda_{1,T}=\lambda$, KSC first estimates the variable selection stability $S(\lambda)$ by randomly partitioning the samples ${(\by_t, \wh\bZ_{t-1}): t=2,\cdots, T}$ into two subsets, repeating this process $B$ times. Then, $\lambda_{1,T}$ is chosen as $\wh \lambda_{1,T}=\min {\lambda: \frac{S(\lambda)}{\max_{\lambda'}S(\lambda')}\geq 1-\alpha_T}$, where we set $B=50$ and $\alpha_T=0.4$.
	In summary, there are three methods for selecting tuning parameters in \textbf{Algorithm 2}, and their impact on the algorithm's performance is discussed in the following sections.
	
}

In our simulations, the coefficient matrices $\bA \in \mathbb{R}^{p_1 \times p_1}$ and $\bB \in \mathbb{R}^{p_2 \times p_2}$ are generated as follows: For a given dimension $p_1$, let $r_1$ be the proportion of nonzero entries in $\bA$. For each row of $\bA$, $\left\lfloor \frac{p_1}{2} \right\rfloor$ entries are generated from $U[1, 2]$, and the remaining $p_1 - \left\lfloor \frac{p_1}{2} \right\rfloor$ entries are generated from $U[-2, -1]$. Next, $p_1 - \left\lfloor r_1 p_1 \right\rfloor$ entries are set to zero, and the elements are randomly rearranged to form one row of $\bA$. Finally, we rescale $\bA$ so that $|\bA|_F = 1$. The procedure for generating $\bB$ follows the same steps as for $\bA$, except that $\bB$ is rescaled to satisfy $\rho (\bA) \rho (\bB) = 0.9$.

To measure the accuracy of \textbf{Algorithm 2} in recovering non-zero elements, we define the following sets for $\bA = (a_{i,j})$ and $\wh \bA = (\wh a_{i,j})$:
\begin{align*}
	&S_1 = \{(i,j) | a_{i,j} = 0, \wh a_{i,j} = 0\}, & S_2 = \{(i,j) | a_{i,j} = 0, \wh a_{i,j} \neq 0\}, \\
	&S_3 = \{(i,j) | a_{i,j} \neq 0, \wh a_{i,j} \neq 0\}, & S_4 = \{(i,j) | a_{i,j} \neq 0, \wh a_{i,j} = 0\}.
\end{align*}
The recovery accuracy for non-zero elements in $\bA$ is then defined as 
\begin{equation}
	\text{cr}(\wh \bA) = \frac{|S_1| + |S_3|}{p_1 \times p_1},
\end{equation}
which represents the proportion of correctly estimated zero and non-zero entries in $\wh \bA$ relative to $\bA$. The recovery accuracy for non-zero elements in $\bB = (b_{i,j})$ is defined similarly.

Table~\ref{tb_lasoo_cr} reports $\text{cr}(\wh \bA )$ and $\text{cr}(\wh \bB )$ under different settings and tuning methods from 100 independent replications. \textbf{Algorithm 2} shows varying performance depending on the tuning method. Specifically, tuning with sdCV results in the highest accuracy for recovering non-zero elements but also the largest error in $\log(|\wh\bS - \bS |_F)$. In contrast, tuning with mCV yields the best accuracy for $\log(|\wh\bS - \bS |_F)$ but the worst recovery accuracy. Finally, tuning with KSC provides a balanced performance, achieving comparable results in both recovery accuracy and the error in $\log(|\wh\bS - \bS |_F)$, making it a well-rounded choice.

\begin{table}
	\centering
	\footnotesize
	\caption{The performance of \textbf{Algorithm 2} under different settings and different tuning methods:  sdCV, mCV and KSC.}
	\label{tb_lasoo_cr}
	\begin{tabular}{@{}cccccccccccc@{}}
		\toprule
		&     &        & \multicolumn{3}{c}{sdCV} & \multicolumn{3}{c}{mCV} & \multicolumn{3}{c}{KSC} \\ \cmidrule(lr){4-6} \cmidrule(lr){7-9} \cmidrule(lr){10-12}
		$(p_1, p_2)$              & $T$   & $E_{\text{alse}}$ & $\text{cr}(\wh \bA )$  & $\text{cr}(\wh \bB )$     & $E_{\text{lasso}}$   & $\text{cr}(\wh \bA )$  & $\text{cr}(\wh \bB )$     & $E_{\text{lasso}}$   &$\text{cr}(\wh \bA )$  & $\text{cr}(\wh \bB )$    & $E_{\text{lasso}}$   \\ \midrule
		\multirow{4}{*}{$(6, 4)$} & 100 & -0.558 & 0.979   & 0.999   & -0.324    & 0.691   & 0.725   & -0.848    & 0.931   & 0.965   & -0.797    \\
		& 200 & -0.906 & 0.987   & 0.999   & -0.579    & 0.682   & 0.688   & -1.211    & 0.930   & 0.951   & -1.116    \\
		& 400 & -1.260 & 0.996   & 1       & -0.751    & 0.706   & 0.716   & -1.573    & 0.929   & 0.962   & -1.525    \\
		& 800 & -1.616 & 0.998   & 1       & -0.963    & 0.703   & 0.716   & -1.920    & 0.936   & 0.966   & -1.856    \\ \midrule
		\multirow{4}{*}{$(9, 6)$} & 100 & -0.247 & 0.93    & 0.983   & 0.001     & 0.663   & 0.674   & -0.495    & 0.914   & 0.930   & -0.287    \\
		& 200 & -0.602 & 0.956   & 0.990   & -0.248    & 0.662   & 0.661   & -0.856    & 0.915   & 0.944   & -0.628    \\
		& 400 & -0.962 & 0.974   & 0.997   & -0.463    & 0.671   & 0.663   & -1.228    & 0.918   & 0.945   & -0.990    \\
		& 800 & -1.305 & 0.981   & 0.999   & -0.668    & 0.66    & 0.661   & -1.558    & 0.913   & 0.937   & -1.340    \\ \bottomrule
	\end{tabular}
\end{table}

\begin{figure}[htp]
	\centering\footnotesize
	\includegraphics[scale=0.3]{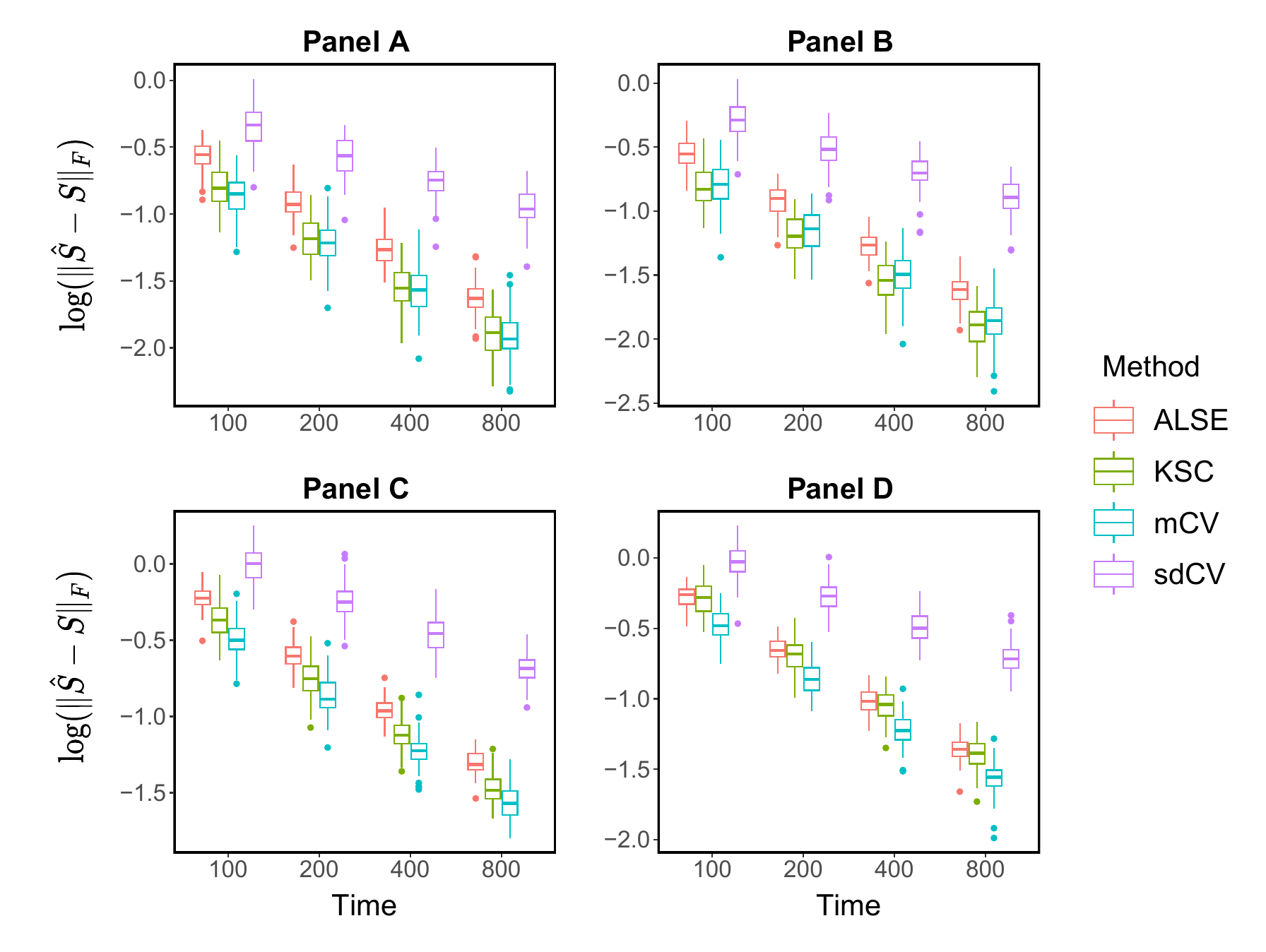}
	\caption{Estimation error $\log(\|\hat{\bS} - \bS \|_F)$ of ALSE and \textbf{Algorithm 2} with different tuning methods (sdCV, mCV and KSC) under different settings of ($p_1, p_2$) and $(r_1, r_2)$.  \textbf{Panel A}: $(p_1, p_2) = (6, 4)$, $(r_1, r_2) = (0.3, 0.3)$; \textbf{Panel B}: $(p_1, p_2) = (6, 4)$, $(r_1, r_2) = (0.4, 0.4)$; \textbf{Panel C}: $(p_1, p_2) = (9, 6)$, $(k_1, k_2) = (0.3, 0.3)$; \textbf{Panel D}: $(p_1, p_2) = (9, 6)$, $(k_1, k_2) = (0.4, 0.4)$. Results are reported from 100 independent repetitions.}
	\label{fig2}
\end{figure}

Next, we compare our estimators with those obtained by the ALSE method from \cite{chenxiaoyang2020} in terms of estimation errors. Figure \ref{fig2} presents the box plot of $ \log(|\wh\bS - \bS |_F) $ from 100 independent replications. Detailed results on the estimation errors of $\wh\bA$, $\wh\bB$, and $\wh\bS$ using the KSC tuning method are reported in Table~\ref{tb_lasso_cong}, while results for the sdCV and mCV tuning methods are shown in Table~S1 and S2, respectively.
Figure~\ref{fig2} and Table~\ref{tb_lasso_cong} indicate that the estimators produced by \textbf{Algorithm 2} generally outperform those from the ALSE method, as the Lasso solutions yield smaller estimation errors in most cases. This suggests that the proposed procedure generates more accurate estimators. Additionally, consistent with previous findings, \textbf{Algorithm 2} tuned with KSC shows comparable performance to that tuned with mCV, and significantly outperforms the sdCV-tuned version, as seen in Figure~\ref{fig2}, Table~\ref{tb_lasso_cong}, Table~S1, and Table~S2. It also performs better than ALSE (Figure~\ref{fig2}).

In summary, \textbf{Algorithm 2} tuned by KSC demonstrates satisfactory performance in both recovery accuracy and estimation error, outperforming ALSE with significantly sparser coefficient matrices that are easier to interpret. Therefore, we recommend it for real data applications. From this point forward, we will refer to \textbf{Algorithm 2} as \textbf{Algorithm 2} tuned by KSC.

\begin{table}[htpb] \footnotesize
	\centering
	\caption{The estimation errors of the estimators obtained by ALSE and \textbf{Algorithm 2} with tuning parameters method KSC.}
	\label{tb_lasso_cong}
	\begin{tabular}{cccccccccc}
		\toprule
		&        & \multicolumn{4}{c}{Algorithm 2 tunned by KSC} & \multicolumn{4}{c}{ALSE}           \\
		\cmidrule(lr){3-6} \cmidrule(lr){7-10}  $(p_1, p_2)$ & $\rho$ & T = 100   & 500      & 1000    & 2000    & T = 100 & 500    & 1000   & 2000   \\ \midrule
		\multicolumn{10}{c}{$ \log(\|\wh \bA - \bA \|_F) $}                                                                                              \\
		$(6, 4)$                                             & 0.5    & -1.603    & -2.173   & -2.553  & -2.96   & -1.465  & -1.903 & -2.243 & -2.597 \\
		$(6, 4)$                                             & 0.9    & -2.364    & -2.853   & -3.225  & -3.571  & -2.196  & -2.645 & -2.993 & -3.344 \\
		$(9, 6)$                                             & 0.5    & -1.753    & -2.308   & -2.710  & -3.127  & -1.546  & -2.003 & -2.355 & -2.712 \\
		$(9, 6)$                                             & 0.9    & -2.634    & -3.124   & -3.494  & -3.893  & -2.334  & -2.793 & -3.144 & -3.495 \\ \midrule
		\multicolumn{10}{c}{$ \log(\|\wh \bB - \bB \|_F) $}                                                                                              \\
		$(6, 4)$                                             & 0.5    & -1.319    & -1.817   & -2.162  & -2.511  & -1.265  & -1.802 & -2.152 & -2.508 \\
		$(6, 4)$                                             & 0.9    & -1.420    & -1.902   & -2.253  & -2.612  & -1.425  & -1.921 & -2.272 & -2.631 \\
		$(9, 6)$                                             & 0.5    & -0.913    & -1.383   & -1.737  & -2.076  & -0.880  & -1.375 & -1.730 & -2.070 \\
		$(9, 6)$                                             & 0.9    & -1.059    & -1.521   & -1.874  & -2.219  & -1.064  & -1.538 & -1.887 & -2.231 \\ \midrule
		\multicolumn{10}{c}{$ \log(\| \wh\bS - \bS \|_F) $}                                                                                              \\
		$(6, 4)$                                             & 0.5    & -0.843    & -1.389   & -1.755  & -2.134  & -0.736  & -1.216 & -1.561 & -1.918 \\
		$(6, 4)$                                             & 0.9    & -0.990    & -1.479   & -1.840  & -2.192  & -0.891  & -1.359 & -1.706 & -2.063 \\
		$(9, 6)$                                             & 0.5    & -0.469    & -0.989   & -1.366  & -1.742  & -0.324  & -0.798 & -1.151 & -1.505 \\
		$(9, 6)$                                             & 0.9    & -0.697    & -1.174   & -1.534  & -1.902  & -0.523  & -0.988 & -1.338 & -1.687 \\ \bottomrule
	\end{tabular}
\end{table}

\subsection{Real Data Examples}\label{sec4.2}

In this section, we apply the proposed regularized estimation methods to two real-world examples. We utilize three iterative algorithms: the ALSE method from \cite{chenxiaoyang2020}, \textbf{Algorithm 1}, and \textbf{Algorithm 2} to estimate the coefficient matrices $\bA$ and $\bB$ in Model (\ref{md}). We then examine the out-of-sample forecasting errors produced by the MAR(1) model using parameters estimated by the three approaches. The empirical findings demonstrate that our proposed algorithms achieve a high degree of sparsity in modeling the matrix-variate data, resulting in a significant reduction in model parameters. Furthermore, our methods outperform the ALSE method in terms of out-of-sample forecast errors.


\subsubsection{Wind Speed Data}

In this example, we apply our methodology to a wind speed dataset consisting of the east–west component of the wind speed vector over a region between latitudes $14^{\circ}$S and $16^{\circ}$N and longitudes $145^{\circ}$E and $175^{\circ}$E in the western Pacific Ocean.
The data records the average wind speed every 6 hours on a 
$17 \times 17$ grid (covering 289 locations) from November 1992 to February 1993, resulting in 
$T = 480$ and $p_1 = p_2 = 17$. Previous studies by \cite{Hsu2012} and \cite{hsu2021} have shown evidence of non-stationary spatial dependence, while indicating temporal stationarity with positive temporal correlations.

\begin{figure}[!htp]
	\centering
	\includegraphics[scale = 0.4]{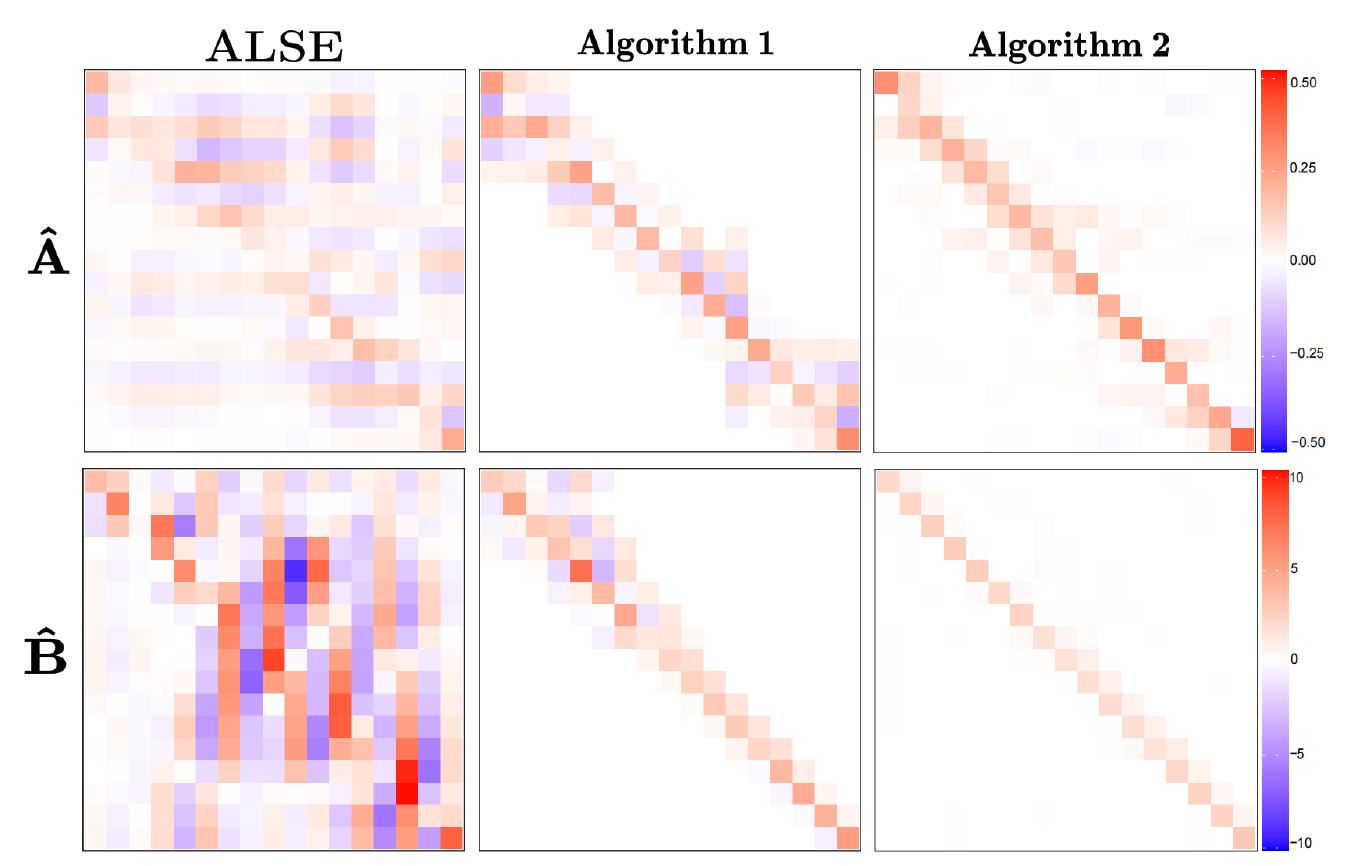}
	\caption{The coefficient matrices estimated by ALSE, \textbf{Algorithm 1}, and \textbf{Algorithm 2} from U-wind dataset. The first row shows results of $\wh \bA$, and the second row shows those of $\wh \bB$ obtained by different methods.}\label{fig:rd1}
\end{figure}

The observed data was divided into training data $ \{ \bY_1, \dots, \bY_{400} \}$ and validation data $\{ \bY_{401}, \dots, \bY_{480} \}$. 
We apply the three iterative estimation methods (ALSE, \textbf{Algorithm 1}, and \textbf{Algorithm 2}) to obtain the estimated coefficients 
$ \wh \bA $ and $ \wh \bB $ in the model (\ref{md1}). To evaluate the performances of these methods, we calculate the average prediction mean-squared error (PMSE).
$$ \text{PMSE} = \frac{1}{289\times 80} \sum_{t = 400}^{479} \| \wh \bY_{t+1} - \bY_{t+1} \|_F, $$
and the average prediction mean-absolute error (PMAE)
$$ \text{PMAE} = \frac{1}{289\times 80} \sum_{t = 400}^{479} \| \wh \bY_{t+1} - \bY_{t+1} \|_1, $$
on the validation data, where $ \wh \bY_{t+1} = \wh \bA \bY_{t} \wh \bB'$. Furthermore, the sparsity of the coefficient matrices estimated by these methods, in terms of the proportions of zero entries in each matrix and the number of iteration steps, is reported in Table \ref{tb:rd1}. From Table~\ref{tb:rd1}, we see that our proposed \textbf{Algorithm 1} under the banded case and \textbf{Algorithm 2} under the sparse case perform better than the ALSE method in terms of both PMSE and PMAE. Moreover, the degree of sparsity of the coefficient matrices estimated by our methods is much higher than that by the ALSE method, which implies that our methods greatly simplify the model. In general, \textbf{Algorithm 2} with $ \lambda_1 = \lambda_2 = 0.1 $ in the Lasso estimation performed best among the three methods. The bandwidths of $ \wh \bA $ and $ \wh \bB $ chosen by the proposed BIC are 4 and 5, respectively. The heat maps of the $ \wh \bA $ and $ \wh \bB $ estimated by the three methods are shown in Figure \ref{fig:rd1}, which clearly illustrate the sparsity of the parameters estimated by our methods.

\begin{table}[htp]
	\centering
	\caption{The performance of the three different methods on  wind speed data.}\label{tb:rd1}
	\begin{tabular}{cccccc}
		\toprule
		Method & PMSE    & PMAE    & Sparsity of $\wh \bA$ & Sparsity of $\wh \bB$ & Iteration step \\ \midrule
		ALSE    & 0.16612 & 0.20007 & 0             & 0             & 45              \\
		\textbf{Algorithm 1} & 0.16563 & 0.19983 & 0.6851        & 0.7578        & 8               \\
		\textbf{Algorithm 2}  & 0.16341 & 0.19665 & 0.7647        & 0.8651        & 6               \\ \bottomrule
	\end{tabular}
\end{table}

\subsubsection{Economic Indicator Data}
The data in this example consists of quarterly observations of four economic indicators: the long-term interest rate (first-order differenced series), GDP growth (first-order log differenced of the GDP series), total manufacturing production growth (first-order log differenced of the production series), and the total consumer price index (growth from the previous period). These indicators are sourced from five countries: Canada, France, Germany, the United Kingdom, and the United States. The dataset spans from the first quarter of 1990 to the fourth quarter of 2016, resulting in a $4 \times 5$ matrix-valued time series with a time length of $T = 107$. This data has been previously studied in \cite{chenxiaoyang2020}, and we follow the same data arrangement order for the five countries and four economic indicators. The data can be obtained from the Organisation for Economic Co-operation and Development (OECD) at \href{https://data.oecd.org/}{https://data.oecd.org/}.

\begin{figure}
	\centering
	\includegraphics[scale = 0.15]{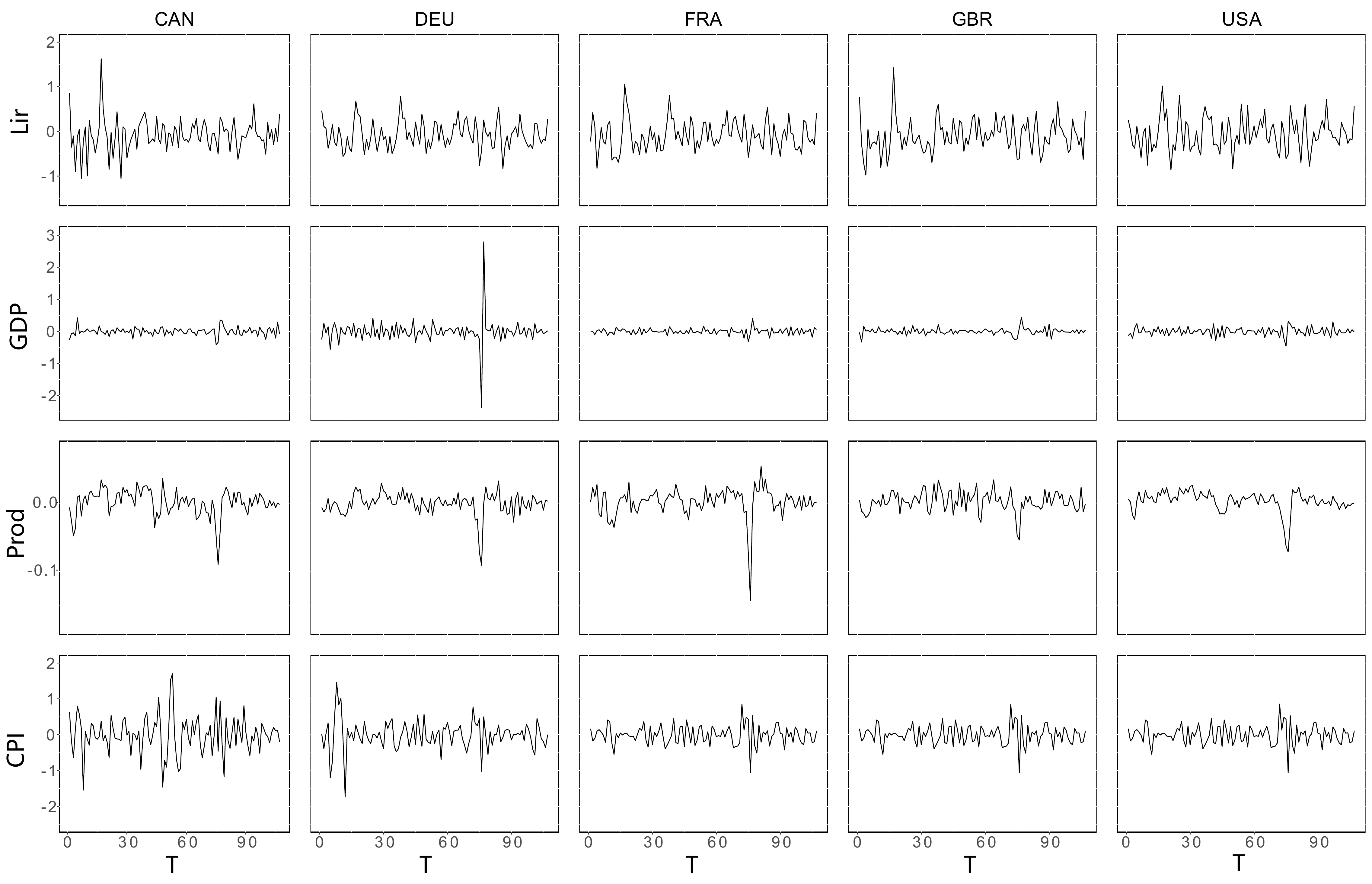}
	\caption{Time series of four indicators: Long-term interest rate(first order difference), GDP growth(log difference), Total manufacture Production growth(log difference), and Total CPI growth from last period subtracted the sample quarterly means from five countries.}\label{fig:rd2}
\end{figure}

To eliminate seasonal effects, we adjusted the seasonality of the Consumer Price Index (CPI) by subtracting the sample quarterly means. In Figure \ref{fig:rd2}, we present these adjusted data, with rows and columns corresponding to different economic indicators and countries.

We compare the out-of-sample rolling forecast performances of the MAR model (\ref{md1}) estimated using the three iterative methods mentioned above. The out-of-sample rolling forecasting begins from $t = 87$ to $t = 106$. For each time point, we fit the model using the data ${ \bY_1, \bY_2, \dots, \bY_t }$ to obtain the estimated coefficient matrices $\wh \bA$ and $\wh \bB$. We then compute the predictive value $ \wh \bY_{t+1} = \wh \bA \bY_t \wh \bB' $, as well as the 1-norm predictive error $ \| \wh \bY_{t+1} - \bY_{t+1} \|_1 $ and the F-norm predictive error $ \|\wh \bY{t+1} - \bY_{t+1} \|_F $. The averages of the 1-norm and F-norm errors from $t = 87$ to $t = 106$ for the three iterative methods are reported in Table \ref{tb5}. Notably, both our Lasso iterative method and the banded iterative method outperform the ALSE method from \cite{chenxiaoyang2020}.

\begin{table}[htp]
	\centering
	\caption{The Out-sample rolling forecast performances of the three methods(ALSE, \textbf{Algorithm 1} and \textbf{Algorithm 2}) on Economic indicator data.}\label{tb5}
	
	\begin{tabular}{cccc}
		\toprule
		& ALSE       & \textbf{Algorithm 1}     & \textbf{Algorithm 2}    \\ \midrule
		MAE & 0.7249358 & 0.7094906 & 0.7027468 \\ 
		MSE & 0.8533359 & 0.8378350 & 0.8236744 \\ \bottomrule
	\end{tabular}
\end{table}

\section{Conclusion}\label{sec5}
In this paper, we studied statistical estimators for high-dimensional matrix-valued autoregressive models under two different settings: when the parameter matrix is banded or sparse. We established the asymptotic properties of these estimators. Both simulations and real data analyses demonstrate the advantages of our new methods over existing ones.  The proposed method can be treated as another option in the toolbox for
modeling high-dimensional matrix-variate time series and the dynamic models can be useful to practitioners
who are interested in out-of-sample forecasting.

\vskip .65cm
\noindent
$^1$Center for Data Science, Zhejiang University. 
E-mail: \{jianghj,12335035,12235025\}@zju.edu.cn
\vskip 2pt

\noindent
$^2$School of Mathematical Sciences, University of Electronic Science and Technology of China. 
E-mail: zhaoxing.gao@uestc.edu.cn\\
\noindent

\end{document}